\documentclass[aps,prl,amsmath,amssymb,
reprint,longbibliography,superscriptaddress,
preprintnumbers]{revtex4-1}
\pdfoutput=1
\usepackage{graphicx}
\usepackage{dcolumn}
\usepackage{bm}
\usepackage[T1]{fontenc} 
\usepackage{amsmath}
\usepackage{amssymb}
\usepackage{amsfonts}
\usepackage{mathtools}
\usepackage{microtype}
\usepackage{tikz}
\usepackage{bm}
\usepackage[utf8]{inputenc}
\usepackage{blkarray}
\usepackage{bigstrut}
\usepackage{nccmath}
\usepackage{enumitem}
\usepackage{braket}
\usepackage{dsfont}
\usepackage{lipsum}
\usepackage{physics}
\usepackage{xcolor}
\usepackage{color}
\usepackage{float}
\usepackage{subfig}
\usepackage{hyperref}
\usepackage{footmisc}
\usepackage{xargs}
\usepackage[colorinlistoftodos,prependcaption,textsize=tiny]{todonotes}
\newcommandx{\will}[2][1=]{\todo[linecolor=red,backgroundcolor=red!25,bordercolor=red,#1]{\underline{\textit{Will:}} #2}}

\usepackage{caption}
\usepackage{booktabs}
\usepackage{slashed}

\usepackage[export]{adjustbox}

\def\una{ \left( \frac{\alpha_{\rm b}}{\pi} \right) }

\def\FeynArts{{{\sc FeynArts}}}
\def\FeynCalc{{{\sc FeynCalc}}}

\def\Mathematica{{{\sc Mathematica}}}

\def\Aida{{{\sc Aida}}}

\allowdisplaybreaks

\begin{document}

\newcommand{\munich}{Max-Planck-Institut f{\"u}r Physik, Werner-Heisenberg-Institut, 80805 M{\"u}nchen, Germany.
}
\newcommand{\valencia}{Instituto de F\'{\i}sica Corpuscular, UVEG--CSIC, 46980 Paterna, Spain.}

\preprint{MPP-2021-84; ZU-TH 29/21}


\title{Two-Loop Four-Fermion Scattering Amplitude in QED}

\author{R.~Bonciani }
\email{roberto.bonciani@roma1.infn.it}
\affiliation{Dipartimento di Fisica, Universit\`a di Roma “La Sapienza” and INFN Sezione di Roma, 00185 Roma, Italy}
\author{A.~Broggio}
\email{alessandro.broggio@unimib.it}
\affiliation{ Universit\`a degli Studi di Milano-Bicocca and INFN Sezione di Milano-Bicocca, Piazza della Scienza 3, I--20126 Milano, Italy}
\author{S.~Di Vita}
\affiliation{INFN, Sezione di Milano, Via Celoria 16, 20133 Milano, Italy}
\affiliation{Dipartimento di Fisica, Università degli Studi di Milano, Via Celoria 16, 20133 Milano, Italy}
\author{A.~Ferroglia}
\email{aferroglia@citytech.cuny.edu}
\affiliation{Physics Department, New York City College of Technology, The City University of New York, 300 Jay Street, Brooklyn, NY 11201, USA}
\affiliation{The Graduate School and University Center, The City University of New York, 365 Fifth Avenue, New York, NY 10016, USA}
\author{M.~K.~Mandal }
\email{manojkumar.mandal@pd.infn.it}
\affiliation{INFN, Sezione di Padova, Via Marzolo 8, 35131 Padova, Italy}
\affiliation{Dipartimento di Fisica e Astronomia, Universit\`a di Padova, Via Marzolo 8, 35131 Padova, Italy}
\author{P.~Mastrolia }
\email{pierpaolo.mastrolia@pd.infn.it}
\affiliation{Dipartimento di Fisica e Astronomia, Universit\`a di Padova, Via Marzolo 8, 35131 Padova, Italy}
\affiliation{INFN, Sezione di Padova, Via Marzolo 8, 35131 Padova, Italy}
\author{L.~Mattiazzi }
\email{mattiazzi@pd.infn.it}
\affiliation{INFN, Sezione di Padova, Via Marzolo 8, 35131 Padova, Italy}
\affiliation{Dipartimento di Fisica e Astronomia, Universit\`a di Padova, Via Marzolo 8, 35131 Padova, Italy}
\author{A.~Primo}
\email{aprimo@physik.uzh.ch}
\affiliation{Department of Physics, University of Z{\"u}rich, CH-8057 Z{\"u}rich, Switzerland}
\author{J.~Ronca}
\email{joron@uv.es}
\affiliation{Dipartimento di Fisica, Universit\`a di Napoli Federico II and INFN, Sezione di Napoli, I-80126 Napoli, Italy}
\author{U.~Schubert}
\email{ulrichsc@buffalo.edu}
\affiliation{Department of Physics,University at Buffalo, The State University of New York, Buffalo 14260, USA}
\author{W.~J. Torres~Bobadilla }
\email{torres@mpp.mpg.de}
\affiliation{\munich}
\author{F.~Tramontano}
\email{francesco.tramontano@unina.it}
\affiliation{Dipartimento di Fisica, Universit\`a di Napoli Federico II and INFN, Sezione di Napoli, I-80126 Napoli, Italy}
\date{\today}

\begin{abstract}
We present the first fully analytic evaluation of the transition amplitude for the scattering of a mass-less into a massive pair of fermions at the two-loop level in Quantum Electrodynamics. 
Our result is an essential ingredient for the determination of the
electromagnetic coupling within scattering reactions,
beyond the currently known accuracy, which has a crucial impact on the
evaluation of the anomalous magnetic moment of the muon. 
It will allow, in particular, for a precise determination of the leading hadronic contribution to the $(g-2)_{\mu}$ in the MUonE experiment at CERN, and therefore can be used to shed light on the current discrepancy between the Standard Model prediction and the experimental measurement for this important physical observable.

\end{abstract}

\maketitle

\noindent{\it Introduction -- }The Muon g-2 collaboration at Fermilab has recently confirmed~\cite{Abi:2021gix} that the observed magnetic activity of the muon is compatible with the earlier findings obtained at Brookhaven National Lab~\cite{Bennett:2002jb,Bennett:2004pv,Bennett:2006fi}. The anomalous magnetic moment of the muon, $(g-2)_\mu$, shows a $4.2\sigma$ deviation from the prediction of the Standard Model of elementary particles (SM)~\cite{Aoyama:2020ynm}.
However, the theoretical determination of this quantity, obtained via dispersive techniques, might be affected by the improper estimation of the hadronic corrections to the muon--photon interaction, which could be responsible of such a discrepancy. Alternative results obtained through lattice QCD  calculations point towards a possible mitigation of the tension between theory and experiments~\cite{Borsanyi:2020mff}.

Recently, a novel experiment, MUonE, has been proposed at CERN, with the goal of measuring the running of the effective electromagnetic coupling at low momentum transfer in the space-like region~\cite{Abbiendi:2016xup}. As proposed in~\cite{Calame:2015fva}, this  measurement would provide an independent determination of the leading hadronic contribution to the $(g-2)_\mu$. Such a measurement relies on the precise determination of the angles of the outgoing particles emerging from the elastic muon-electron scattering~\cite{Abbiendi:2016xup,loimuone,Abbiendi:2019qtw,Abbiendi:2021xsh}.
To extract the running of the effective electromagnetic coupling from the experimental data, the pure perturbative electromagnetic contribution to the electron-muon cross section must be controlled at least up to the second order in the fine-structure constant~\cite{Banerjee:2020tdt}.

The scattering of a muon $\mu$ off an electron $e$ in Quantum Electrodynamics~(QED) is the simplest reaction among fundamental leptons of different flavors, and represents a paradigmatic case of charged particles interaction  mediated by a neutral gauge boson. The Leading Order~(LO) process is known since the mid 1950's~\cite{Berestetskii:1956aaa}, 
while the Next-to-Leading Order~(NLO) radiative corrections
were computed in~\cite{Nikishov:1961aaa,Eriksson:1961,Eriksson:1963,VanNieuwenhuizen:1971yn,Kukhto:1987uj,Bardin:1997nc,Kaiser:2010zz}, and more recently studied in~\cite{Alacevich:2018vez}. 
The two-loop diagrams contributing to the Next-to-Next-to-leading order~(NNLO) virtual corrections were evaluated in~\cite{Bern:2000ie} 
assuming purely massless fermions. 
At the energies of the MUonE experiment, the muon mass plays an important role for the description of the radiative pattern and cannot be neglected~\cite{Banerjee:2020tdt}.
Nevertheless, the evaluation of Feynman integrals usually becomes more demanding as the number of massive particles present either in the loops or in the external states increases. 

NNLO QED corrections involve the two-loop amplitude along with the real-virtual and the double-real emission terms. 
While the matrix elements for the last two contributions can be  calculated without difficulties using standard techniques, their
integration over the corresponding phase spaces is complicated by the presence of infrared (soft and collinear) singularities,
as well as the presence of masses in both the initial and final state of the scattering process.
In order to obtain predictions for fully differential observables, it is necessary to adopt a subtraction procedure.
The Abelian nature of the interaction leads us to believe that the computational techniques already used for other processes at the LHC can be successfully adapted to this purpose~\cite{Heinrich:2020ybq,TorresBobadilla:2020ekr}. 
Preliminary Monte Carlo simulations
for $\mu e$ scattering have already been performed by including parts of the NNLO corrections \cite{CarloniCalame:2020yoz,Banerjee:2020rww}. These simulations account for a subset of the two-loop graphs, not yet including the four-point diagrams with complete dependence on the lepton masses.
The complete two-loop amplitude is then a missing crucial ingredient for the computation of the full NNLO QED corrections.

In this work, we present the first fully analytic evaluation of the renormalized two-loop amplitude for four fermion scattering in QED, 
$f^- + f^+ + F^- + F^+ \to 0$, with $f$ and $F$ representing a massless and a massive lepton respectively.
In the past years, we have developed efficient mathematical techniques for the evaluation of multi-loop integrals in dimensional regularization, such as the {\it adaptive integrand decomposition}~\cite{Mastrolia:2016dhn,Mastrolia:2016czu,Mastrolia:2019aid} and the {\it Magnus exponential method} for differential equations \cite{Argeri:2014qva,DiVita:2014pza}. The combination of these techniques with the more traditional decomposition through {\it integration-by-parts identities}~(IBPs)~\cite{Chetyrkin:1981qh,Laporta:2001dd}, allowed us to obtain for the first time a complete analytic formula for the renormalized two-loop amplitude of a $2 \to 2$ process with a non-vanishing mass in internal and external lines.

The one- and two-loop amplitudes presented in this Letter can be applied, for instance, to the case where the light fermion is an electron, $f = e$, and the heavy fermion is a muon, $F=\mu$, and can be used in the elastic scattering $e \mu \to e \mu$, as well as in crossing related processes, such $e^+e^- \to \mu^+ \mu^-$.
If the elastic scattering is the key process of the MUonE experiment, the muon pair production in $e^+e^-$ annihilation is a key process for the center-of-mass energy calibration at present and future $e^+e^-$ colliders, such as BESIII~\cite{Ablikim:2017duf}, 
BELLE-II~\cite{Kou:2018nap},
CEPC~\cite{Smiljanic:2021bbb}, and FCCee~\cite{Blondel:2019jmp}. Therefore, a precise knowledge
of the radiative effects would improve the precision of the results obtainable at these machines.

The structure of the infrared (IR) singularities of the massless and massive gauge theory scattering amplitudes has been studied in ~\cite{Catani:1998bh,Sterman:2002qn,Aybat:2006mz,Aybat:2006wq,Gardi:2009qi,Gardi:2009zv,Becher:2009cu,Becher:2019avh,Becher:2009qa,Penin:2005kf,Penin:2005eh,Mitov:2006xs,Becher:2007cu,Bonciani:2007eh,Bonciani:2008ep,Becher:2009kw}.
In this work, the determination of the virtual NNLO corrections is complemented by the investigation of the IR singularities of scattering amplitudes in QED, which involve massive particles, and whose universal structure can be determined within Soft Collinear Effective Theory (SCET), along the lines of the method presented in \cite{Becher:2009qa, Becher:2009kw}.
The agreement of the residual IR poles coming from the direct diagrammatic calculus
of the renormalized amplitude
with the IR poles predicted within SCET is an important validation of the diagrammatic calculation. We explicitly verify this agreement in the case of $f^-f^+ \to F^- F^+$ process.

Additionally, let us observe that the two-loop diagrams considered here, also appear in the (color stripped) Abelian subset of graphs contributing to heavy-quark pair production in Quantum Chromodynamics (QCD)~ \cite{Czakon:2008zk, Bonciani:2008az, Bonciani:2009nb, Baernreuther:2013caa,Badger:2021owl}.
The similarities of the analytic structure of the two-loop amplitude between $q {\bar q} \to t {\bar t}$ in QCD and $f^-f^+ \to F^- F^+$ in QED, where $q$ and $f$ are treated as massless, is exploited to test the structure
of the singularities coming from QED diagrams through a tuned comparison to the Abelian part of known results in QCD.

Recently, the evaluation of integrals coming from planar diagrams~\cite{Fael:2019nsf,Fael:2018dmz,Heller:2021gun} indicates that the computation of four-fermion scattering  
amplitudes at two loops in QED, by keeping full dependence on the masses of all the involved leptons, might become the subject of near-future investigation. \\


\noindent{\it Scattering Amplitude --}
\label{sec:Amplitude}
\begin{figure}[t]
\includegraphics[scale=0.9]{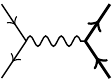}\;
\vspace{0.2cm}

\includegraphics[scale=0.8]{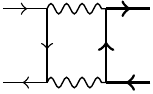}\quad
\includegraphics[scale=0.8]{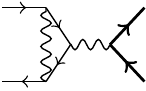}\quad
\includegraphics[scale=0.9]{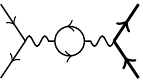}
\vspace{0.2cm}

\includegraphics[scale=0.8]{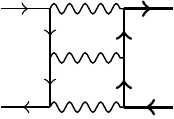}\quad
\includegraphics[scale=0.8]{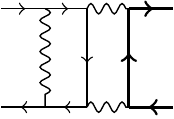}\quad
\includegraphics[scale=0.8]{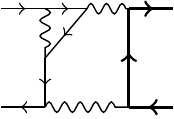}\quad
\includegraphics[scale=0.8]{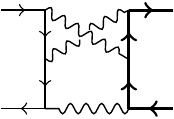}

\caption{Representative diagrams for the process $f^- f^+ \to F^- F^+$:
tree-level (top), one-loop graphs (middle), 
two-loop graphs (bottom).
Thin lines indicate a lepton $f$ while thick lines indicate a lepton $F$. Wavy lines are photons.
}
\label{fig:diagrams}
\end{figure}
We consider the four-fermion scattering process involving a mass-less and a massive lepton pair, 
\begin{equation}
	f^{-}(p_1) + f^{+}(p_2) \rightarrow F^{-}(p_3) + F^{+}(p_4)\; ,
\end{equation}
with $m_f=0$ and $m_F = M \ne 0$.
The Mandelstam invariants, defined as
$	s=(p_1+p_2)^2, \, \, t=(p_1-p_3)^2, $ and $ \, u=(p_2-p_3)^2 $,
satisfy the condition $s + t + u = 2 M^2$. 

The four-point bare amplitude ${\cal A}_{\rm b}$
admits a perturbative expansion in the bare coupling constant $\alpha_{\rm b} \equiv e_{\rm b}^2/4 \pi$, which, up to the inclusion of the second-order corrections,
reads
\begin{align}
   \label{eq:unrenormamp}
	{\cal A}_{\rm b}\left(\alpha_{\rm b}\right)
	=  
	4 \pi &\alpha_{\rm b}\, S_{\epsilon} \, \mu^{-2 \epsilon}\notag \\
	& \times 
	\bigg[ 
	{\cal A}_{\rm b}^{(0)}
	+ \una {\cal A}_{\rm b}^{(1)} + \una^2 \!\! {\cal A}_{\rm b}^{(2)} 
	\bigg] \,, 
\end{align} 
where ${\cal A}_{\rm b}^{(n)}$
indicates the $n$-loop bare amplitude, 
$S_{\epsilon} \equiv (4\pi e^{-\gamma_{E}})^{\epsilon}$
and $\mu$ is the 't Hooft mass scale.
The Leading Order (LO) term 
${\cal A}_{\rm b}^{(0)}$, 
referred to as {\it Born term},
receives contribution from a single tree-level Feynman diagram, shown in the upper row of Fig.~\ref{fig:diagrams}. 
The squared LO amplitude, summed over the final spins and averaged over the initial states, reads,
\begin{align}
{\cal M}^{(0)}_{\rm b} &=
\frac{1}{4}\sum_{\rm spins} 
|{\cal A}^{(0)}_{\rm b}|^2 \nonumber \\
   &= 
   \frac{1}{s^2}\big[2(1-\epsilon) s^2 + 4 \left(t-M^2\right)^2+4 s t \big]
   \, ,
\label{eq:bareborn}   
\end{align}
%
for external states treated in $d=4-2\epsilon$ space-time dimensions
according to the conventional dimensional regularization (CDR)
scheme~\cite{Gnendiger:2017pys}, that we use throughout the whole computation. The interferences of one- and two-loop bare amplitudes 
with the Born amplitude read
\begin{equation}
{\cal M}^{(n)}_{\rm b} = 
\frac{1}{4}\sum_{\rm spins} 2\, \text{Re} ( {\cal 
A}^{(0)*}_{\rm b} \, 
{\cal A}^{(n)}_{\rm b} ) \,, \ \text{for $n=1,2$}
\, .
\label{eq:bareinterference}
\end{equation} 
\noindent{\it Analytic Evaluation -- }
The analytic evaluation of $\mathcal{M}_{\text{b}}^{\left(1\right)}$
and $\mathcal{M}_{\text{b}}^{\left(2\right)}$ is completely automated, within an in-house software, which can be applied to generic one- and two-loop amplitudes.
The \Mathematica{}
package \FeynArts~\cite{Hahn:2000kx}
is used to generate Feynman diagrams contributing to the
one- and two-loop corrections to the scattering amplitudes 
as well as the counter-term diagrams required for the renormalization: 6 diagrams and 3 counter-term diagrams at one loop; 69 diagrams (12 of which vanish because of Furry’s theorem) and 55 counter-term diagrams at two loops.
%
Representative one- and two-loop diagrams are shown in the second and third row of Fig.~\ref{fig:diagrams}, respectively.
The spin sums and the algebraic manipulation to simplify the Dirac-$\gamma$ algebra are carried out by means of the \FeynCalc~\cite{Mertig:1990an,Shtabovenko:2016sxi,Shtabovenko:2020gxv} package.
Each $n$-loop graph $G$ (interfered with the Born amplitude) corresponds to an integrand written in terms of scalar products between external, $p_i^\nu$, and internal, $k_i^\nu$, momenta. Therefore, Eq.\eqref{eq:bareinterference} can be generically written as,
\begin{equation}
{\cal M}^{(n)}_{\rm b} = 
( S_{\epsilon} )^n
 \int \prod_{i=1}^n \frac{d^d k_i}{(2 \pi)^d} \, 
\sum_{G} \frac{N_G}{\prod_{\sigma \in G} D_\sigma } \quad , 
\end{equation}
where:
$N_G=N_G (p_i,k_i)$ indicates the numerator, and 
$D_\sigma = D_\sigma(p_i,k_i,M)$ are the denominators corresponding to the internal lines of $G$.

Integrands are simplified by employing the {\it adaptive integrand decomposition method}, implemented in the 
\Aida{} framework~\cite{Mastrolia:2019aid}. 
The intermediate results emerging from the integrand decomposition can be further simplified by means of the IBP identities~\cite{Chetyrkin:1981qh,Laporta:2001dd}.
Our software is interfaced with the  publicly available codes {\sc Reduze}~\cite{vonManteuffel:2012np} 
and {\sc Kira}~\cite{Maierhoefer:2017hyi}, 
and, for each diagram, it produces the files for the automated generation of the IBP relations.
After the decomposition phase, 
the interference terms  $\mathcal{M}_{\text{b}}^{\left(n\right)}$
are written as linear combination 
of a set of independent integrals, 
say ${\bf I}^{(n)}$,
called {\it master integrals} (MIs), 
\begin{eqnarray}
    {\cal M}^{(n)}_{\rm b}  &=& 
    {\mathbb C}^{(n)} \, \cdot \,
    {\bf I}^{(n)} \, ,
\end{eqnarray}
where ${\mathbb C}^{(n)}$ is a vector of coefficients, depending on $\epsilon$ and  the kinematic variables, $s,t,M^2$. 
In particular,
$\mathcal{M}_{\text{b}}^{(1)}$
and $\mathcal{M}_{\text{b}}^{(2)}$ are conveniently expressed, 
in terms of 12 and 264 MIs, respectively,  
analytically computed: 
two- and three-point functions have been known since long~\cite{Gehrmann:1999as,Bonciani:2003te,Bonciani:2003hc},
while planar and non-planar four-point integrals were computed
in~\cite{DiVita:2018nnh,Mastrolia:2017pfy},
using the differential equation method via Magnus exponential, and independently in \cite{Bonciani:2008az,Bonciani:2009nb,Becchetti:2019tjy}.
The analytic expressions of 
${\cal M}^{(n)}_{\rm b}$ can 
be written as a Laurent series around 
$d=4$ space-time dimensions ($\epsilon =0$), 
with coefficients that contain Generalized Polylogarithms (GPLs)~\cite{Goncharov:1998kja}, defined as iterated integrals, through the recursive formula
\begin{eqnarray}
  G(w_n,\ldots,w_1 ; \tau) 
  &\equiv
  \int_0^\tau \frac{dt}{t-w_n} 
  G(w_{n-1},\ldots,w_1 ;t) \, , \quad
\end{eqnarray}
with 
$G(w_{1};t) \equiv \log(1-t/w_1)$.
The arguments $w_i$ are known as {\it letters},
and their number, corresponding to the number of nested integrations, is called {\it weight}.
The two-loop interference term contains 
4063 GPLs with up to weight four, whose arguments are written in terms of 18 letters, $w_i=w_i(x,y,z)$, which depend on the Mandelstam variables through the relations,
$- t/M^2 = x \, ,$ 
$- s/M^2 = (1-y)^2/y \, ,$
$- (u - M^2)/(t-M^2) = z^2/y \, $ 
(see~\cite{DiVita:2018nnh,Mastrolia:2017pfy} for more details). \\

\begin{figure}[t]
\begin{minipage}{1.5cm}
\includegraphics[scale=0.8]{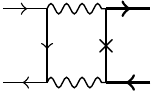}\qquad
\end{minipage}
\quad
\begin{minipage}{1.5cm}
\includegraphics[scale=0.8]{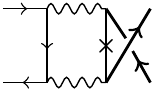}\qquad 
\end{minipage}
\quad
$2 \times$\!\!\!\!
\begin{minipage}{1.5cm}
\includegraphics[scale=0.8]{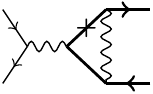}\qquad
\end{minipage}
\quad
$2 \times$\!\!\!\!
\begin{minipage}{1.5cm}
\includegraphics[scale=0.9]{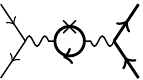}
\end{minipage}

\caption{Diagrams for mass renormalization. The ${\boldsymbol{\times}}$ symbol represents the insertion of a mass counter-term. }
\label{fig:ctdiagrams}
\end{figure}

\noindent{\it Renormalization --}
The one- and two-loop diagrams contributing to $\mathcal{M}_{\text{b}}^{(1)}$ and 
$\mathcal{M}_{\text{b}}^{(2)}$ 
contain infrared (IR) and ultraviolet (UV) divergences.
To remove the UV divergences,
the bare lepton fields ($\psi_\ell$, with 
$\ell=f,F$, for massless and massive leptons, respectively) and  photon field ($A^\mu$), as well as the bare mass of the massive lepton are renormalized as follows, 
\begin{eqnarray}\label{qedren}
	{\psi}_{\rm b} &=& \sqrt{Z_2} \, {\psi}, \quad
	 A^\mu_{\rm b} = \sqrt{Z_3} \, A^\mu, \quad 
	 M_{\rm b} = Z_M M \ ,
\end{eqnarray}
where, to simplify the notation, the label $\ell$ 
in the lepton fields is understood and restored when required.
The renormalization of the QED interaction vertex,
\begin{eqnarray}
	{\cal L}_{\rm int} = e_{\rm b} \, \bar{\psi}_{\rm b} \, \slashed A_{\rm b} \, \psi_{\rm b} = e \, Z_{1} \, \bar{\psi} \, \slashed A \, \psi \ ,
\end{eqnarray}
can then be entirely fixed using the QED Ward identity, that implies $Z_1 = Z_2$. In particular, this leads to a simple relation between the renormalized charge and the bare charge (obtained by applying Eq.~\eqref{qedren} to the bare interaction term and comparing the two renormalized expressions)
$	e \, Z_1 = e_{\rm b} \, Z_2 \, \sqrt{Z_3}$, 
therefore, one has 
$e = e_{\rm b} \sqrt{Z_3}$.
The lepton wave functions and the mass of the massive lepton are renormalized in the on-shell scheme, namely, 
$Z_{2,f} = Z_{2,f}^{\scriptscriptstyle\rm OS}$,   
$Z_{2,F} = Z_{2,F}^{\scriptscriptstyle\rm OS}$,   
$Z_{M} = Z_{M}^{\scriptscriptstyle\rm OS}$.  
The coupling constant is renormalized in 
the $\overline{\text{MS}}$ scheme at the scale $\mu^2$, 
\begin{equation}
  \alpha_{{\rm b}}\,S_{\epsilon}=\alpha(\mu^{2})\,\mu^{2\epsilon}\,Z_{\alpha}^{\scriptscriptstyle\overline{\text{MS}}}\,,  
\end{equation}
with $Z_\alpha^{\scriptscriptstyle\overline{\text{MS}}} = 1/Z_3^{\scriptscriptstyle\overline{\text{MS}}}$.
The renormalized amplitude is obtained 
by multiplying the bare amplitude with a factor $\sqrt{Z_{2,\ell}}$ for 
any external lepton $\ell$, hence,
\begin{eqnarray}
	\label{eq:renamp}
	{\cal A} 
	&=& Z_{2,f} \, Z_{2,F} \, 
	\hat{{\cal A}}_{\rm b} \ ,
\end{eqnarray}
where $\hat{{\cal A}}_{\rm b} = {\cal A}_{\rm b}
\left(
\alpha_{\rm b} = \alpha_{\rm b}(\alpha), 
M_{\rm b} = M_{\rm b}(M)
\right)
	$,
namely expressing the bare coupling and mass in terms of their renormalized counterparts.
Let us observe that ${\cal A}$ 
depends on four renormalization constants,
namely 
$Z_\alpha^{\scriptscriptstyle\overline{\text{MS}}},Z_{2,f}^{\scriptscriptstyle\rm OS},Z_{2,F}^{\scriptscriptstyle\rm OS},Z_{M}^{\scriptscriptstyle\rm OS}$.
To simplify the notation in the following, these are simply indicated as 
$Z_{j}$, with $j = \{\alpha, f, F, M\}$, respectively.
The renormalization constants admit a perturbative expansions in $\alpha$, 
\begin{align}
\label{eq:Zexpansion}
Z_{j} &= 1 + \left(\frac{\alpha}{\pi}\right) \delta Z_{j}^{(1)} + \left(\frac{\alpha}{\pi}\right)^2 \delta Z_{j}^{(2)} 
+ {\cal O}(\alpha^3) \ ,
\end{align}
and their expressions can be obtained (either directly or after abelianization) from \cite{Broadhurst:1991fy,Melnikov:2000zc,Czakon:2007ej,Baernreuther:2013caa}.
After substituting in Eq.~\eqref{eq:renamp} the expansions of the bare amplitude,
given in Eq.~\eqref{eq:unrenormamp}, and the ones of the renormalization constants, given in Eq.~\eqref{eq:Zexpansion},
the UV renormalized two-loop amplitude reads 
\begin{eqnarray}
   \label{eq:twoloopR}
	{\cal A} \left(\alpha \right)  
	&=&  
	4 \pi \alpha
	\bigg[
	{\cal A}^{(0)}
	+ \left(\frac{\alpha}{\pi}\right) {\cal A}^{(1)} 
	+ \left(\frac{\alpha}{\pi}\right)^2 {\cal A}^{(2)} 
	\bigg] \ ,
\end{eqnarray}
up to  second order corrections in $\alpha$.
The $n$-loop coefficients ${\cal A}^{(n)}$ are given in terms of the ones appearing in the bare amplitude as
\begin{subequations}
\begin{eqnarray}
	{\cal A}^{(0)} &=& {\cal A}_{\rm b}^{(0)},
	\\
	{\cal A}^{(1)}  &=& {\cal A}_{\rm b}^{(1)} 
	+
	\Big( \delta Z_{\alpha}^{(1)}
	+ \delta Z_{F}^{(1)} \Big) {\cal A}_{\rm b}^{(0)},
	\\
	{\cal A}^{(2)} &=& {\cal A}_{\rm b}^{(2)}
	+ \Big(2 \delta Z_{\alpha}^{(1)} + \delta Z_{F}^{(1)} \Big) {\cal A}_{\rm b}^{(1)} 
	\nonumber \\
	&+& \Big(\delta Z_{\alpha}^{(2)} 
	+ \delta Z_{F}^{(2)} + \delta Z_{f}^{(2)} 
	+ \delta Z_{F}^{(1)} \delta Z_{\alpha}^{(1)}
	\Big) {\cal A}_{\rm b}^{(0)} \nonumber \\
	&+& \delta Z_M^{(1)} 
	{\cal A}_{\rm b}^{(1, \text{mass CT})} \label{eq:aux1}
\, . 
\end{eqnarray}
	\label{eq:RenAmpInBareAmp}
\end{subequations}
The last term in Eq.~(\ref{eq:aux1}) contains the extra contribution of one-loop diagrams having an insertion of the mass counter-term in the massive propagators in all possible ways, as depicted in Fig.~\ref{fig:ctdiagrams}. \\
The bare coupling $\alpha_{\rm b}$ and the bare amplitudes  
${\cal A}^{(n)}_{\rm b}$ ($n=0,1,2$),
appearing in Eqs.~\eqref{eq:bareborn} and 
(\ref{eq:bareinterference}),
can be replaced by the corresponding renormalized quantities
$\alpha$ and 
${\cal A}^{(n)}_{\rm}$, 
to build the Born term, ${\cal M}^{(0)}$, 
and the renormalized interference terms, at one loop,
${\cal M}^{(1)}$, and at two loops, ${\cal M}^{(2)}$.
The latter two quantities constitute the
main results of this Letter. \\

\noindent{\it Infrared Structure --}
The IR poles appearing in the two-loop corrections after UV renormalization can independently be obtained starting from the tree-level and the one-loop amplitudes, by following the same procedure employed to study the infrared structure of QCD amplitudes \cite{Becher:2009qa, Becher:2009kw}.  

The structure of the IR poles is governed by an anomalous dimension $\Gamma$ that has the following structure,
\begin{align} \label{eq:gamma}
    \Gamma =&   \gamma_{\text{cusp}}\left(\alpha \right) \ln \left( -\frac{s}{\mu^2} \right) 
    + 2 \gamma_{\text{cusp}}\left(\alpha \right)\ln \left( \frac{t - M^2}{u -M^2} \right) \nonumber \\
    &+ \gamma_{\text{cusp,M}}\left(\alpha, s \right) + 2 \gamma_h\left(\alpha \right) + 2 \gamma_\psi\left(\alpha \right)\, ,
\end{align}
where the $\gamma_i$ ($i \in \{\text{cusp};\text{cusp,M};h;\psi\}$) coefficients  up to ${\mathcal O}(\alpha^2)$ are extracted in analogy to the QCD case \cite{Becher:2009kw,Becher:2009qa,Hill:2016gdf}.
We compute the analytic expression of the two-loop amplitude ${\cal M}^{(2)}$ for the process $f^- f^+ \to F^- F^+$ both in the non-physical region $s<0, t<0$ as well as directly in the production region.
In this physical region, the imaginary part of the anomalous dimension in Eq.~\eqref{eq:gamma} is computed by adding an infinitesimal positive imaginary part to $s$.
One can then introduce the IR renormalization factor $Z_{\text{IR}}$,
\begin{align} \label{eq:logZ}
 \ln Z_{\text{IR}} &= \frac{\alpha}{4 \pi} \left( \frac{\Gamma'_0}{4 \epsilon^2} 
 + \frac{\Gamma_0}{2 \epsilon} \right) 
 \nonumber \\ & 
 + \left(\frac{\alpha}{4 \pi}\right)^2
 \Biggl( - \frac{3 \beta_0 \Gamma'_0}{16 \epsilon^3}  
 + \frac{\Gamma'_1 - 4 \beta_0 \Gamma_0}{16 \epsilon^2} 
 + \frac{\Gamma_1}{4 \epsilon} \Biggr) 
 \nonumber \\ & 
 + \mathcal{O}\left( \alpha^3 \right) \, ,
 \end{align}
where $\Gamma_i, \Gamma'_i$ and $\beta_i$ are the coefficients of the expansion of $\Gamma$, its derivative w.r.t. $\ln \mu$, and the QED beta function, respectively. 
The IR poles of the $n^{\rm th}$-order term ${\mathcal M}^{(n)}$ can be calculated using $Z_{\text{IR}}$ and the lower order contributions, ${\mathcal M}^{(0)}, \hdots, {\mathcal M}^{(n-1)}$.
In particular, the IR pole structures at one and two loops are found to be, 
\begin{subequations}
\begin{eqnarray}
\label{eq:predpoles1L}
   \left.{\mathcal M}^{(1)} \right|_{\text{poles}} &=& 
   \frac{1}{2} Z_{1}^\text{IR} \left.{\mathcal M}^{(0)}
  \right|_{\text{poles}} \, , \\
\label{eq:predpoles2L}
     \left.{\mathcal M}^{(2)} \right|_{\text{poles}} &=& 
     \frac{1}{8} \Big[\!\!
     \left( Z_{2}^{\text{IR}} - \left(Z_{1}^{\text{IR}}\right)^2 \right) 
   {\mathcal M}^{(0)} \nonumber \\ 
   && \qquad \qquad 
   + 2 \, Z_{1}^{\text{IR}} \, {\mathcal M}^{(1)}
   \left. \Big] \right|_{\text{poles}} \, .
    \qquad \quad
\end{eqnarray} 
\label{eq:predpoles}
\end{subequations}
All functions ${\mathcal M}^{(n)}$ in the r.h.s. of Eqs.~\eqref{eq:predpoles} must be evaluated in $d=4 -2 \epsilon$ space-time dimensions.
The factors $Z_i^{\text{IR}}$ are the coefficients of the series expansion of $Z_{\text{IR}}$ in powers of $\alpha/(4 \pi)$.

The IR poles structure in Eqs.~(\ref{eq:predpoles}), reconstructed starting from the tree-level and one-loop amplitudes, is in perfect
agreement with the one obtained starting from Eq.~(\ref{eq:aux1}) and directly calculating the two-loop diagrams. This provides a non trivial test of the complete two-loop calculation. \\

\begin{figure}[t]
\centering
\includegraphics[scale=0.45]{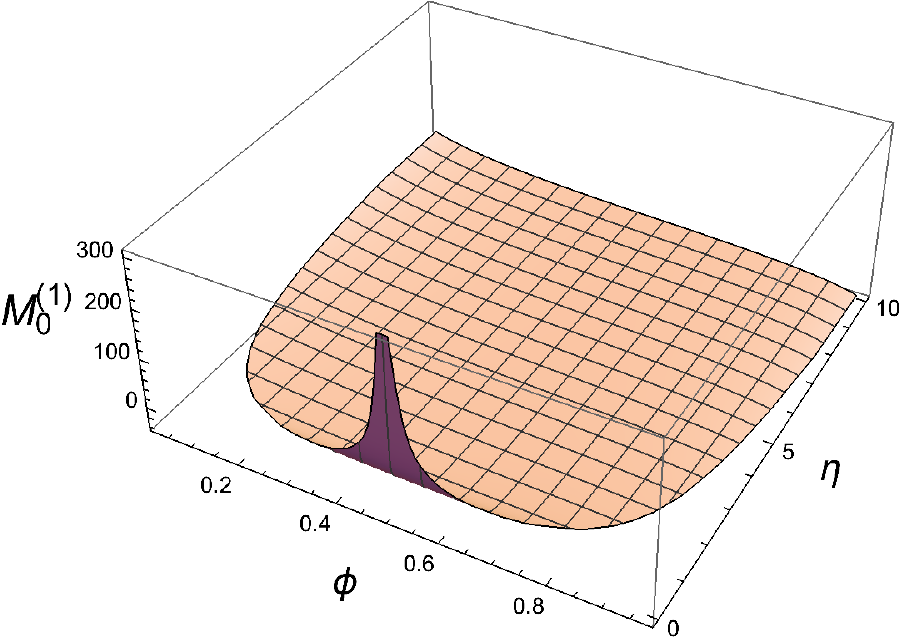}
\includegraphics[scale=0.45]{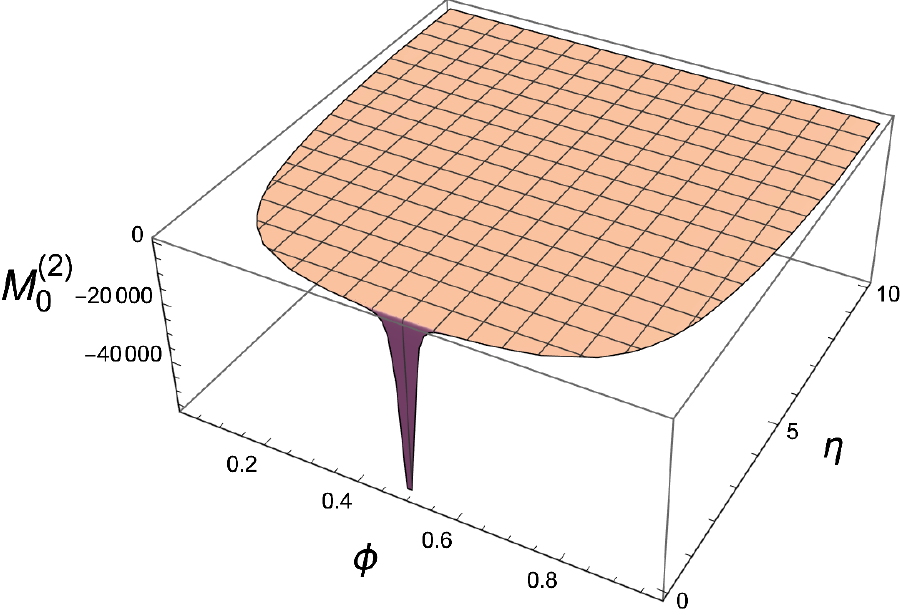}
\caption{Three-dimensional plots of the finite terms
${\cal M}_0^{(i)}$, $i=1,2$
of the renormalized one- and two-loop amplitudes,  in 
Eqs.~\eqref{eqs:finalamplitude1L}, \eqref{eqs:finalamplitude2L}, where $n_l=1$, $n_h=1$.  \\
Here, $\eta = s/(4M^2) - 1$, $\phi = -(t-M^2)/s$.
}
\label{fig:3d_plots_total1L2L}
\end{figure}

\noindent{\it Results --}
The analytic results of the interference contributions 
${\cal M}^{(1)}$ and ${\cal M}^{(2)}$ are given  
as Laurent series in $\epsilon$
\begin{subequations}
\begin{eqnarray}
\label{eqs:finalamplitude1L}
 {\cal M}^{(1)} &=& 
 \frac{{\cal M}^{(1)}_{-2}}{\epsilon^2} \, 
 \!+\! 
 \frac{{\cal M}^{(1)}_{-1}}{\epsilon} \, 
 \!+\!
 { {\cal M}^{(1)}_{0} } 
 \!+\!
 { {\cal M}^{(1)}_{1} \epsilon} 
 +
 {\cal O}(\epsilon^2)
 \, , \quad \\
\label{eqs:finalamplitude2L}
 {\cal M}^{(2)} &=& 
 \frac{ {\cal M}^{(2)}_{-4}}{\epsilon^4} \, 
 + \ldots +
 \frac{ {\cal M}^{(2)}_{-1}}{\epsilon} \, 
 +
 { {\cal M}^{(2)}_{0} } 
 + 
 {\cal O}(\epsilon)\ .
 \end{eqnarray}
\end{subequations}
The analytical expression of ${\cal M}^{(1)}$ is computed both in the non-physical region, and in the pair production region, $s > 4M^2$, $t < 0$. The latter is required to predict the IR poles of ${\cal M}^{(2)}$ directly in the production region; the analytical expression of ${\cal M}^{(2)}$ is computed in the non-physical region, $s < 0$, $t < 0$, and its analytic continuation is performed numerically.
%
The renormalized one- and two-loop interference terms are conveniently decomposed in gauge-invariant components, labeled by the number of massless ($n_l$) and massive ($n_h$) closed fermion loops
\begin{subequations}
\begin{eqnarray}
	\label{eq:Amp1LTotCoeff} 
 	{\cal M}^{(1)}
 	&=& A^{(1)} + n_l\,B_{l}^{(1)} + n_h\,C_{h}^{(1)}\,, \\
 	\label{eq:Amp2LTotCoeff}
	{\cal M}^{(2)}
	&=& A^{(2)} + n_l\,B_{l}^{(2)} + n_h\,C_{h}^{(2)} + n_l^2 D_{l}^{(2)} \nonumber \\
	&& + n_h \, n_l \, E_{hl}^{(2)} + n_h^2\,F_{h}^{(2)}\,. 
\end{eqnarray}
\label{eq:AmpTotCoeff}
\end{subequations}
In Fig.~\ref{fig:3d_plots_total1L2L}, we plot the finite part of one- and two-loop renormalized amplitudes ${\cal M}_0^{(i)}$, $i=1,2$ in the physical region. The threshold singularity is clearly visible and well reproduced up to 
very small c.m.e., showing full control of the numerical stability.
The complete formula for the analytic expression of the renormalized two-loop amplitude is rather large ($\sim \! \!60\,$MB) and cannot be reported here. The figures are obtained by evaluating this formula with high precision on $10,\!500$ evenly spaced grid points,
by employing {\sc HandyG}~\cite{Naterop:2019xaf} and {\sc Ginac}~\cite{Vollinga:2004sn} (via the package {\sc PolyLogtools}~\cite{Duhr:2019tlz}) for the numerical evaluation of GPLs. Each evaluation required from seconds CPU time in the almost flat region to up about $1,\!500\,$s CPU time for the configurations approaching the threshold singularity. These grids are available from the authors upon request. \\


\noindent
{\it Other tests --} The master integrals for the Abelian diagrams in QED can be employed to construct the analytic expressions of some gauge-invariant contributions to the 
two-loop amplitude of the process $q {\bar q} \to t {\bar t}$ in QCD~\cite{Czakon:2008zk, Bonciani:2008az, Bonciani:2009nb, Baernreuther:2013caa}: in particular, our results (evaluated in the region of heavy-lepton pair production, and properly accounting for the color factors) agree with the numerical coefficients $E_l^{q}, E_h^{q}, F_l^{q}, F_{lh}^{q}, F_h^{q}$ provided in the Table~1 of Ref.~\cite{Czakon:2008zk, Bonciani:2008az, Baernreuther:2013caa}, which receive contributions from Abelian diagrams only; 
the agreement on the poles of the above mentioned color coefficients, at 
other phase-space points, has been verified using the formula for the IR poles of two-loop amplitudes in QCD, given in Ref.~\cite{Ferroglia:2009ii}. \\

\noindent{\it Conclusion --} We presented the first fully analytic evaluation of the amplitude for the scattering of four fermions in Quantum Electrodynamics, involving two different types of leptons, one of which is treated as massless, up to the second order corrections in the electromagnetic coupling constant.
The calculations were carried out within the dimensional regularization scheme,
and the infrared pole structure of the renormalized amplitude is found to obey the universal behaviour predicted by the Soft Collinear Effective Theory. 
Our result constitutes the first example of a complete scattering amplitude 
for $2 \to 2$ processes, with massless and massive particles in the loops as well as in the external states, involving planar and non-planar diagrams at two loops,  analytically evaluated. 

Our analytic results can be directly applied to the study, at NNLO accuracy, of massive lepton pair production in massless lepton annihilation, and, upon analytic continuation, to the study of the elastic scattering of massive and massless fermions in QED and QCD. 
Notably, the virtual corrections presented here are relevant for the recently proposed 
MUonE  experiment at CERN. 
This experiment is devoted to extraction of the hadronic contribution to the $(g-2)_\mu$ from 
the $\mu e$ scattering. The MUonE  experiment analysis relies on the knowledge of the pure NNLO QED correction to the 
$\mu e$ scattering process, which will be the subject of a dedicated study in the near future.
\\

\noindent
{\it Notes -- } Interested readers can find the expressions of the UV renormalization constants and of
the IR renormalization factor used throughout this Letter, 
and additional plots for the individual contributions of the coefficients $A,B,\hdots,F$ of Eq.~\eqref{eq:AmpTotCoeff} in the Supplemental Material. \\

\noindent
{\it Acknowledgments -- } 
We are indebted to Massimo Passera for insightful discussions, as well as for encouragement at all stages, and for comments on the manuscript. 
It is a pleasure to acknowledge the whole MUonE collaboration for motivating discussions, and for providing a stimulating scientific environment.
In particular, we thank Matteo Fael and Massimo Passera for checks on the $n_h$-corrections.
We also thank Carlo Carloni Calame, Lance Dixon, Federico Gasparotto, Thomas Gehrmann, Stefano Laporta, Giovanni Ossola, Paride Paradisi, Fulvio Piccinini, Vajravelu Ravindran, Germ\'an Rodrigo and Adrian Signer, for interesting discussions at various stages. 
The work of R.B. is partly supported by the italian Ministero della Universit\`a e della Ricerca (MIUR) under grant PRIN 20172LNEEZ.
The work of A.B. is supported by the ERC Starting Grant REINVENT-714788. 
The work of A.F. is supported in part by the PSC-CUNY Award 62243-00 50.
The work of M.K.M. is supported by Fellini - Fellowship for Innovation at INFN funded by the European Union’s Horizon 2020 research and innovation programme under the Marie Sk{\l}odowska-Curie grant agreement No 754496.
The work of A.P. was supported by the Swiss National Science Foundation under grant number 200020-175595.
J.R. and F.T. acknowledge support from INFN.
U.S. is supported by the National Science Foundation awards PHY-1719690 and PHY-1652066.
%
This work is supported by the COST Action CA16201 PARTICLEFACE.

\section{Supplemental Material}

In this supplemental material, we provide further details on the renormalization constants to perform the UV renormalization, and the IR renormalization factor for the predictions of the IR poles of the one- and two-loop four-fermion scattering amplitude in QED, $f^- + f^+ + F^- + F^+ \to 0$, with $f$ and $F$, representing massless and a massive leptons, respectively. \\

\noindent{\it Renormalization Constants --}
%
The renormalization constants for the wave functions of the massive and massless leptons as well as the mass renormalization constant of the massive lepton  admit perturbative expansions in $\alpha$, which can be taken 
 (either directly or after abelianization) from Refs.~\cite{Broadhurst:1991fy,Melnikov:2000zc,Czakon:2007ej,Baernreuther:2013caa}, 
 and read as,
\\
\begin{eqnarray}
Z_{F} &=& 1 + \Bigg(\frac{\alpha}{\pi}\Bigg) \delta Z_F^{(1)} +
\Bigg(\frac{\alpha}{\pi}\Bigg)^2 \delta Z_F^{(2)}\, ,
\\
Z_f &=& 1 + 
 \Bigg(\frac{\alpha}{\pi}\Bigg)^2 \delta Z_f^{(2)}\, ,
\\
Z_M &=& 1 + 
 \Bigg(\frac{\alpha}{\pi}\Bigg) \delta Z_M^{(1)}\, ,
\end{eqnarray}
where the individual perturbative coefficients in the on-shell scheme read as,
\begin{eqnarray}
\delta Z_F^{(1)} &=& 
-\frac{3}{4\epsilon }
-\frac{3 L_{\mu }}{4}
-1
 \nonumber \\
&+&\epsilon  
\left(
-\frac{3 L_{\mu}^2}{8}
-L_{\mu }
-\frac{\pi ^2}{16}
-2
\right)
\\
&+&
\epsilon ^2 
\left(
-\frac{L_{\mu}^3}{8}
-\frac{L_{\mu }^2}{2}
-\frac{\pi ^2 L_{\mu }}{16}
-2 L_{\mu }
+\frac{\zeta_3}{4}
-\frac{\pi^2}{12}
-4
\right)\, ,
\nonumber
\\
\delta Z_F^{(2)} &=& 
n_h\Bigg( 
\frac{1}{\epsilon} \left(\frac{L_{\mu }}{4}+\frac{1}{16}\right)
+
\frac{3 L_{\mu }^2}{8}+\frac{11 L_{\mu }}{24}-\frac{5 \pi
   ^2}{16}+\frac{947}{288}
\Bigg)
\nonumber \\
&+&
n_l
\Bigg(
-\frac{1}{8 \epsilon^2}
+\frac{11}{48 \epsilon }
+\frac{L_{\mu }^2}{8}
+\frac{19 L_{\mu }}{24}
+\frac{\pi ^2}{12}
+\frac{113}{96}
\Bigg)
\nonumber \\
&+& \Bigg(
\frac{9}{32 \epsilon^2}
+\frac{9 L_{\mu}}{16 \epsilon }
+\frac{51}{64 \epsilon }
+\frac{9 L_{\mu }^2}{16}
+\frac{51 L_{\mu }}{32}
-\frac{3 \zeta_3}{2}
\nonumber \\
&-&\frac{49 \pi^2}{64}
+\frac{433}{128}
+\pi ^2 \ln (2)
\Bigg)\, ,
\\
\delta Z_f^{(2)} &=& n_h \Bigg(
\frac{L_{\mu }}{8}+\frac{1}{16 \epsilon }-\frac{5}{96}
\Bigg)\, ,
\\
\delta Z_M^{(1)} &=& \delta Z_F^{(1)} \, ,
\end{eqnarray}
with $L_\mu \equiv \ln ({\mu^2/M^2}) $.

Additionally, the renormalization constant for the electromagnetic coupling up to second order in the $\overline{\text{MS}}$ scheme reads, 
\begin{eqnarray}
Z_{\alpha} &=& 1 + 
 \Bigg(\frac{\alpha}{\pi}\Bigg) 
\Bigg[
\frac{(n_h+n_l)}{3 \epsilon }
\Bigg] 
\nonumber \\
&+&
 \Bigg(\frac{\alpha}{\pi}\Bigg)^2 
 \Bigg[
\frac{\left(n_h+n_l\right){}^2}{9 \epsilon ^2}+\frac{(n_h+n_l)}{8 \epsilon }
\Bigg]\, .
\end{eqnarray}
\\

\noindent{\it Anomalous dimensions -- \label{sec:appB}}
The structure of the IR poles is governed by an anomalous dimension $\Gamma$ whose structure reads as,
\begin{align} 
\label{suppl:eq:gamma}
    \Gamma =&   \gamma_{\text{cusp}}\left(\alpha \right) \ln \left( -\frac{s}{\mu^2} \right) 
    + 2 \gamma_{\text{cusp}}\left(\alpha \right)\ln \left( \frac{t - M^2}{u -M^2} \right) \nonumber \\
    &+ \gamma_{\text{cusp,M}}\left(\alpha, s \right) + 2 \gamma_h\left(\alpha \right) + 2 \gamma_\psi\left(\alpha \right)\, ,
\end{align}
and by
the IR renormalization factor $Z_{\text{IR}}$, defined by exponentiation of the following expression,
\begin{align} 
\label{suppl:eq:logZ}
 \ln Z_{\text{IR}} &= \frac{\alpha}{4 \pi} \left( \frac{\Gamma'_0}{4 \epsilon^2} 
 + \frac{\Gamma_0}{2 \epsilon} \right) 
 \nonumber \\ & 
 + \left(\frac{\alpha}{4 \pi}\right)^2
 \Biggl( - \frac{3 \beta_0 \Gamma'_0}{16 \epsilon^3}  
 + \frac{\Gamma'_1 - 4 \beta_0 \Gamma_0}{16 \epsilon^2} 
 + \frac{\Gamma_1}{4 \epsilon} \Biggr) 
 \nonumber \\ & 
 + \mathcal{O}\left( \alpha^3 \right) \, ,
 \end{align}
where $\Gamma_i, \Gamma'_i$ and $\beta_i$ are the coefficient of the expansion of $\Gamma$, its derivative w.r.t. $\ln \mu$, and the QED beta function, respectively. 
We hereby present the coefficients appearing in the above formulas, up to the needed order in $\alpha$:

\begin{align}
    \gamma_i (\alpha) =  \gamma_0^i \left( \frac{\alpha}{4 \pi} \right) +  \gamma_1^i \left( \frac{\alpha}{4 \pi} \right)^2 +    {\mathcal O} \left(\alpha^3 \right)\, .
\end{align}
The cusp anomalous dimensions for massless leptons have the  coefficients~\cite{Hill:2016gdf},
\begin{equation}
    \gamma_0^{\text{cusp}} = 4 \, , \qquad
    \gamma_1^{\text{cusp}} = - \frac{80}{9} n_l \, ,
\end{equation}
whereas, for massive leptons,
\begin{align}
    \gamma_0^{\text{cusp,M}}(s) = & 
     -\gamma_0^{\text{cusp}} \frac{1+{\bar x}^2}{1-{\bar x}^2} \ln{(-{\bar x})} \, , \nonumber \\
     & \\ 
    \gamma_1^{\text{cusp,M}}(s) = & 
     -\gamma_1^{\text{cusp}} \frac{1+{\bar x}^2}{1-{\bar x}^2} \ln{(-{\bar x})} \, , \nonumber 
    \end{align}
with ${\bar x}$ defined through the relation,
\begin{equation}
    s = M^2 \frac{(1+{\bar x})^2}{{\bar x}} \, .
\end{equation}
The coefficients of the factors related to the 
massless and massive leptons are~\cite{Hill:2016gdf}, 
\begin{align}
    \gamma_0^\psi =& -3 \, , \nonumber \\
    \gamma_1^\psi =& -\frac{3}{2}  +2 \pi^2 -24 \zeta_3 + n_l \left( \frac{130}{27} + \frac{2}{3} \pi^2\right) \,  , \nonumber \\
    \gamma_0^h =& -2 \, , \nonumber \\
    \gamma_1^h =& \frac{40}{9} n_l \, .
\end{align}
The QED beta function has the expansion
\begin{equation}
    \beta \left(\alpha \right) = -2 \alpha \left(\beta_0 + \beta_1 \frac{\alpha}{4 \pi} + {\mathcal O} \left(\alpha^3 \right) \right) \, ,
\end{equation}
in which the only needed coefficient for the present calculation is $\beta_0$,
\begin{equation}
    \beta_0 = - \frac{4}{3} n_l \, .
\end{equation}
The quantity $\Gamma'$  appearing in Eq.~(\ref{suppl:eq:logZ}) is defined as,
\begin{equation}
    \Gamma' \left( \alpha \right) \equiv \frac{\partial}{\partial \ln \mu} \Gamma \left( \alpha \right) = \sum_{n = 0}^\infty \Gamma'_i \left(\frac{\alpha}{4 \pi} \right)^{n +1} \, , 
\end{equation}
with the relevant coefficients,
\begin{align}
    \Gamma'_0 =& - 2 \gamma_0^{\text{cusp}} \, , \nonumber \\
    \Gamma'_1 =& - 2 \gamma_1^{\text{cusp}} \, .
\end{align}

Lastly, the expansion of the renormalization factor $Z_{\text{IR}}$ in Eq.~(\ref{suppl:eq:logZ}) is decomposed as, 
\begin{equation}
    Z_{\text{IR}} =  1+ Z_1^{\text{IR}} \left(\frac{\alpha}{4 \pi} \right) +  Z_2^{\text{IR}} \left(\frac{\alpha}{4 \pi} \right)^2 + {\mathcal O} \left(\alpha^3 \right) \, .
\end{equation}
with, 
\begin{align}
   Z_1^{\text{IR}} =& \frac{\Gamma'_0}{4 \epsilon^2} + \frac{\Gamma_0}{2 \epsilon} \, , \nonumber \\
   & \\
   Z_2^{\text{IR}} =&  \frac{\left(\Gamma'_0\right)^2}{32 \epsilon^4} + \frac{\Gamma'_0}{8 \epsilon^3} \left(\Gamma_0 -\frac{3}{2} \beta_0 \right) +\frac{\Gamma_0}{8 \epsilon^2} \left(\Gamma_0 -2 \beta_0 \right)
   \nonumber \\
   &+\frac{\Gamma'_1}{16 \epsilon^2} + \frac{\Gamma_1}{4 \epsilon} \, .
   \nonumber 
\end{align}

Furthermore, in order to implement the inverse decoupling transformation for the massive leptons, in such a way that one works with $n_l+n_h$ active leptons,
one needs to include an additional term proportional to $n_h$
 in 
$Z_2^{\text{IR}}$:
\begin{equation}
    Z_2^{\text{IR}} \rightarrow 
    Z_2^{\text{IR}} + \delta Z_2^{\text{IR}} \ , 
\end{equation}
where,
\begin{align}
    \delta Z_2^{\text{IR}} \equiv &
    - \frac{2}{3} n_h \biggl[ \frac{\Gamma'_0 }{2 \epsilon^2} 
    L_\mu
    +\frac{\Gamma'_0}{4 \epsilon} 
    \left( 
    L_\mu^2
    + \frac{\pi^2}{6} \right) 
    + \frac{\Gamma_0}{\epsilon} 
    L_\mu
    \Biggr] \, . \nonumber 
\end{align}

\begin{figure}[t]
\centering
\includegraphics[scale=0.45]{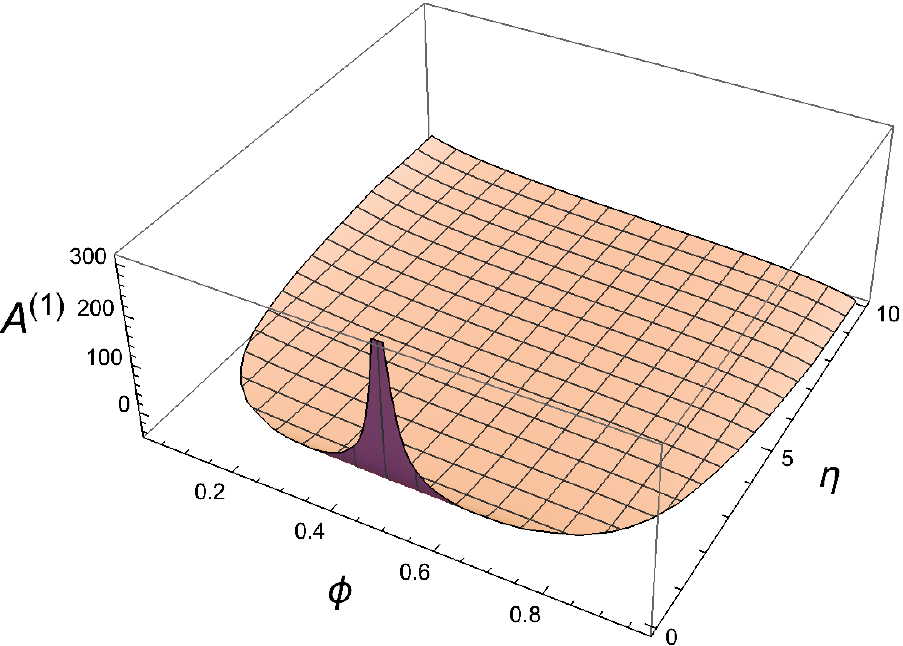}
\includegraphics[scale=0.45]{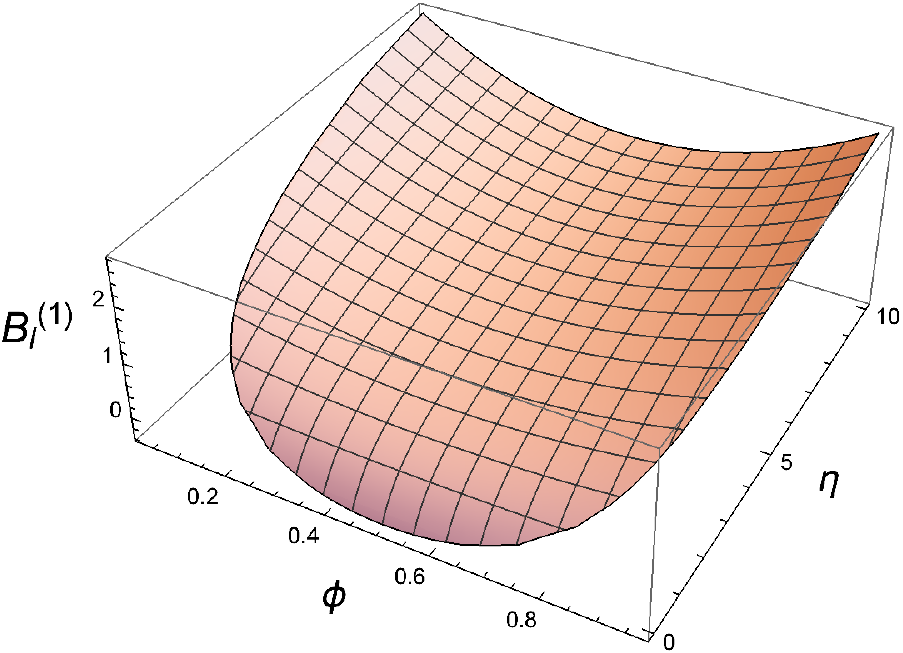}
\includegraphics[scale=0.45]{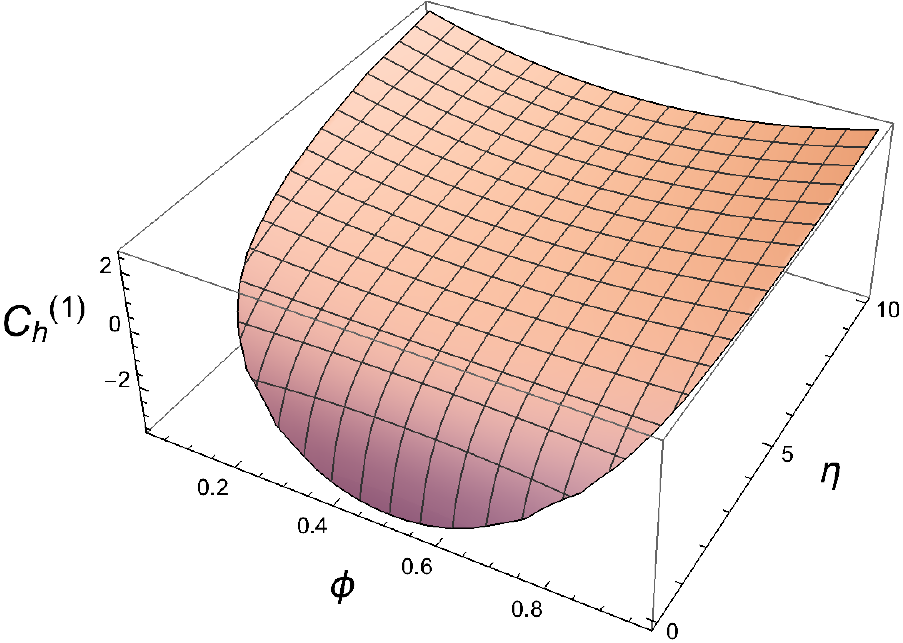}
\caption{Three-dimensional plots of the coefficients (finite part)
appearing in the decomposition of the renormalized one-loop amplitude in Eq.~\eqref{eq:Amp1LTotCoeff}.\\
Here, $\eta = s/(4M^2) - 1$, $\phi = -(t-M^2)/s$.}
\label{fig:3d_plots_1L}
\end{figure}

\begin{figure}[t]
\centering
\includegraphics[scale=0.45]{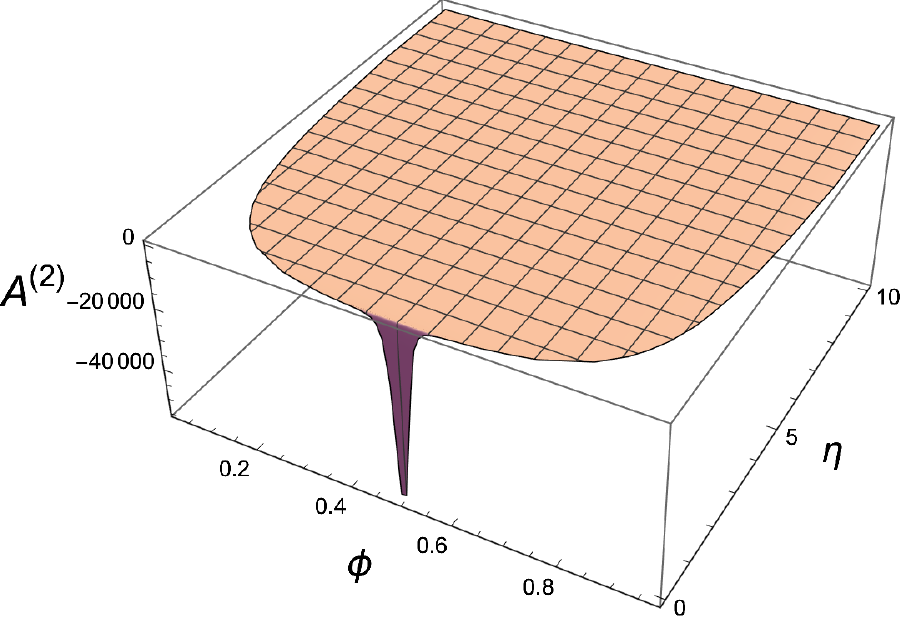}
\includegraphics[scale=0.45]{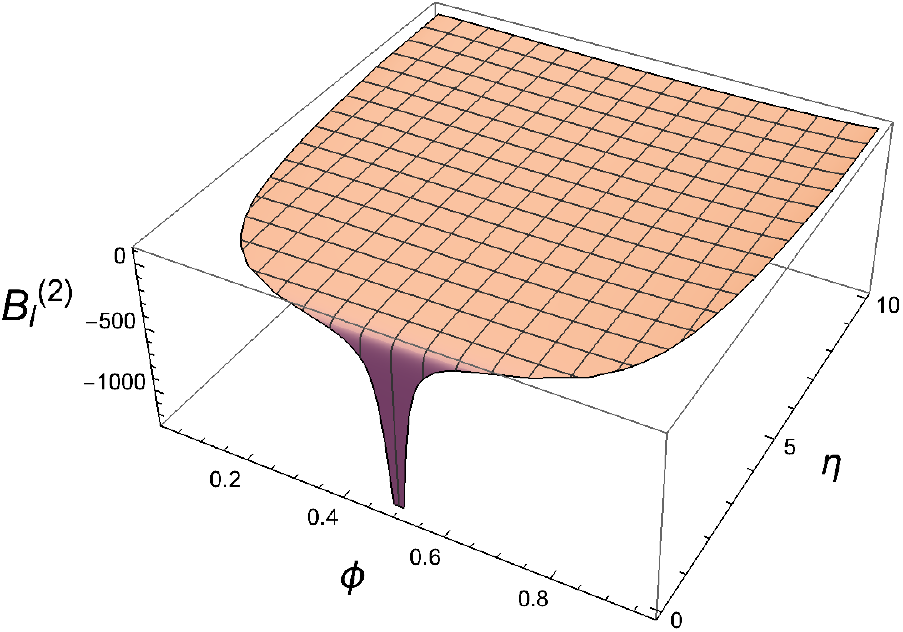}
\includegraphics[scale=0.45]{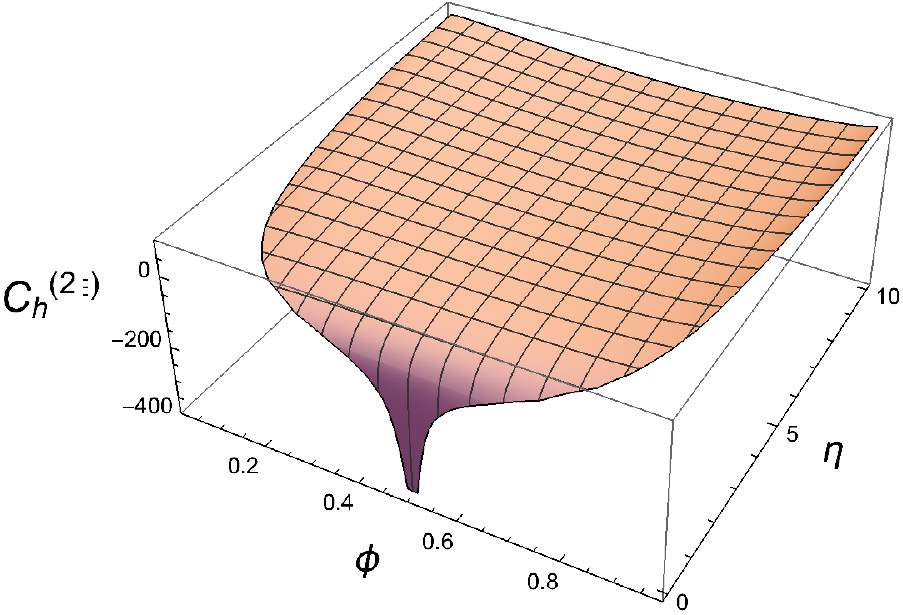}
\includegraphics[scale=0.45]{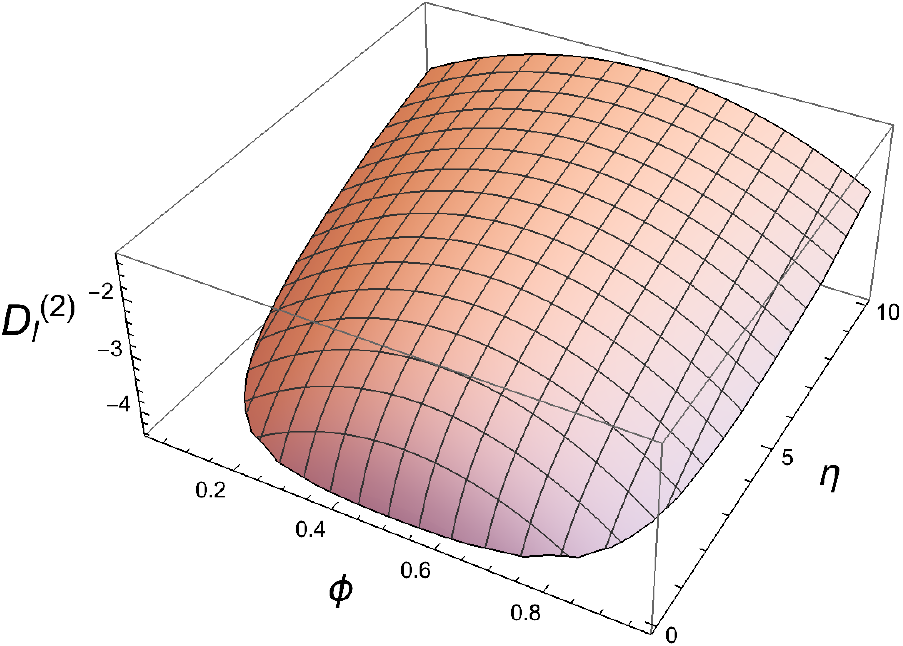}
\includegraphics[scale=0.45]{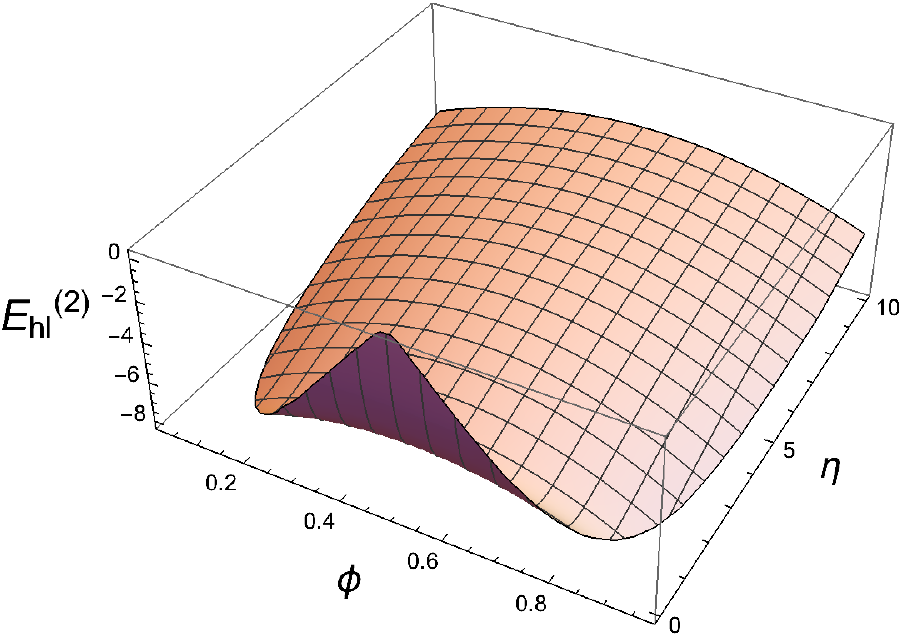}
\includegraphics[scale=0.45]{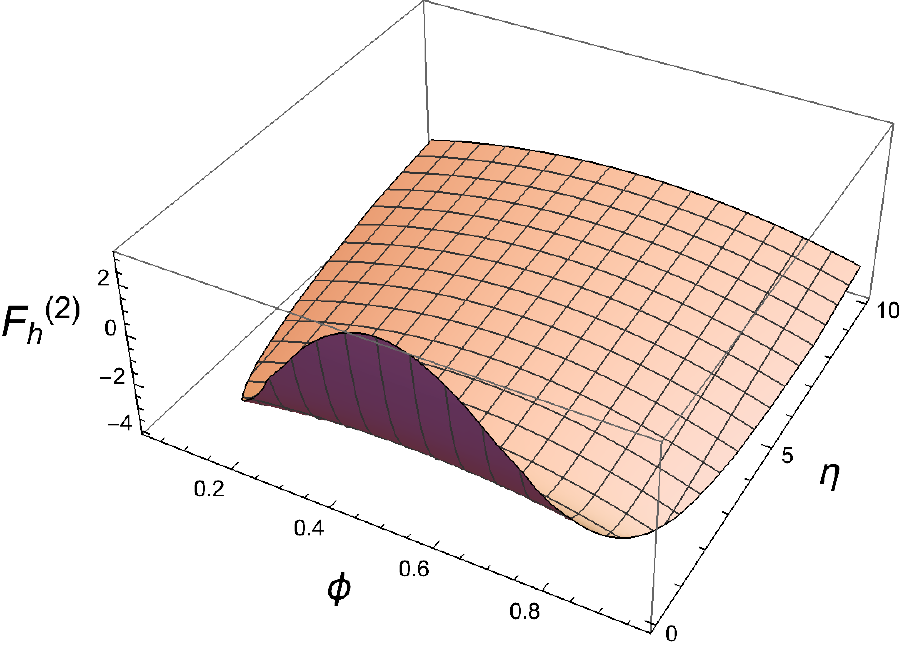}
\caption{Three-dimensional plots of the coefficients (finite part)
appearing in the decomposition of the renormalized two-loop amplitude in Eq.~\eqref{eq:Amp2LTotCoeff}.\\
Here, $\eta = s/(4M^2) - 1$, $\phi = -(t-M^2)/s$.
}
\label{fig:3d_plots_2L}
\end{figure}

\begin{table*}[t!]
\begin{center}    
\begin{tabular}{cp{8em}<{\centering}p{8em}<{\centering}p{8em}<{\centering}p{8em}<{\centering}p{8em}<{\centering}p{8em}<{\centering}} 
    \toprule
    {} & {$\epsilon^{-4}$} & {$\epsilon^{-3}$} & {$\epsilon^{-2}$} & {$\epsilon^{-1}$} & {$\epsilon^{0}$} & {$\epsilon$}\\ 
    \midrule
    ${\cal M}^{(0)}$ & - & - & - & - & $\frac{181}{100}$ & $-2$\tabularnewline
    \midrule
    $A^{(1)}$ & - & - & $-\frac{181}{100}$ & $1.99877525$ & $22.0079572$ & $-11.7311017$\tabularnewline
    $B_{l}^{(1)}$ & - & - & - & - & $-0.069056030$ & $4.94328573$\tabularnewline
    $C_{h}^{(1)}$ & - & - & - & - & $-2.24934027$ & $2.54943566$\tabularnewline
    \midrule
    $A^{(2)}$ & $\frac{181}{400}$ & $-0.499387626$ & $-35.4922919$ & $19.4997261$ & 
    $49.0559119$
    & - \tabularnewline
    $B_{l}^{(2)}$ & - & $-\frac{181}{400}$ & $0.785712779$ & $-16.1576674$ & $-3.75247701$ & -\tabularnewline
    $C_{h}^{(2)}$ & - & - & $1.12467013$ & $-9.50785825$ & $-25.8771503$ & -\tabularnewline
    $D_{l}^{(2)}$ & - & - & - & - & $-3.96845688$ & -\tabularnewline
    $E_{hl}^{(2)}$ & - & - & - & - & $-4.88512563$ & -\tabularnewline
    $F_{h}^{(2)}$ & - & - & - & - & $-0.158490810$ & -\tabularnewline 
    \bottomrule
\end{tabular}
\end{center}
    \caption{Numerical values of the 
    leading order squared amplitude, in Eq.~\eqref{eq:bareborn}, and of the 
    coefficients appearing in the decomposition of the renormalized one- and two-loop amplitudes in Eqs.~\eqref{eq:AmpTotCoeff},
    evaluated at the phase space point $s/M^2= 5$, $t/M^2=-5/4$, $\mu=M$.
    }
    \label{tab:valuesofloopamp}
\end{table*}

\noindent{\it Amplitudes --}
The Born term is shown in Eq.(3) of the Letter, reading as,
\begin{eqnarray}
{\cal M}^{(0)}_{\rm b} &=&
   \frac{1}{s^2}\big[2(1-\epsilon) s^2 + 4 \left(t-M^2\right)^2+4 s t \big]
   \, .
\label{eq:bareborn}   
\end{eqnarray}
The results of the renormalized one- and the two-loop 
interference terms can be found in the
Eqs.(15a,15b) of the Letter, 
as Laurent series in $\epsilon$, hereby reported for conveninece,
\begin{subequations}
\begin{eqnarray}
\label{eqs:finalamplitude1L}
 {\cal M}^{(1)} &=& 
 \frac{{\cal M}^{(1)}_{-2}}{\epsilon^2} \, 
 \!+\! 
 \frac{{\cal M}^{(1)}_{-1}}{\epsilon} \, 
 \!+\!
 { {\cal M}^{(1)}_{0} } 
 \!+\!
 { {\cal M}^{(1)}_{1} \epsilon} 
 +
 {\cal O}(\epsilon^2)
 \, , \quad \\
\label{eqs:finalamplitude2L}
 {\cal M}^{(2)} &=& 
 \frac{ {\cal M}^{(2)}_{-4}}{\epsilon^4} \, 
 + \ldots +
 \frac{ {\cal M}^{(2)}_{-1}}{\epsilon} \, 
 +
 { {\cal M}^{(2)}_{0} } 
 + 
 {\cal O}(\epsilon)\ .
 \end{eqnarray}
\end{subequations}
Each term can be
conveniently decomposed in gauge-invariant components,
see Eqs.(16a,16b), as follows,
\begin{subequations}
\begin{eqnarray}
	\label{eq:Amp1LTotCoeff} 
 	{\cal M}^{(1)}
 	&=& A^{(1)} + n_l\,B_{l}^{(1)} + n_h\,C_{h}^{(1)}\,, \\
 	\label{eq:Amp2LTotCoeff}
	{\cal M}^{(2)}
	&=& A^{(2)} + n_l\,B_{l}^{(2)} + n_h\,C_{h}^{(2)} + n_l^2 D_{l}^{(2)} \nonumber \\
	&& + n_h \, n_l \, E_{hl}^{(2)} + n_h^2\,F_{h}^{(2)}\,. 
\end{eqnarray}
\label{eq:AmpTotCoeff}
\end{subequations}
In Figs 1 and 2, we plot the finite parts of the individual form factors appearing in the decomposition of the one- and two-loop amplitudes, given in Eqs.~(\ref{eq:AmpTotCoeff}).
Finally, in Table \ref{tab:valuesofloopamp}, we showcase the numerical values of the coefficients $A,B,C,D,E,F$
in the massive-fermion pair production region
at a particular phase-space point.
Note that, upon accounting for a different 
definition of the Mandelstam variables, this 
benchmark point corresponds to the one used in 
Table~1 of~\cite{Baernreuther:2013caa}.

\bibliographystyle{apsrev}
\bibliography{refs}

\end{document}